\documentclass[11pt]{article}

\usepackage[latin1]{inputenc}
\usepackage[T1]{fontenc}

\newtheorem{theorem}{Theorem}

\newtheorem{definition}{Definition}

\newtheorem{corollary}{Corollary}

\newtheorem{prop}{Proposition}

\newcommand{\FF}[1]{{GF(#1)}}

\newcommand{\Rg}{\mbox{Rk}}

\begin{document}

\title{Properties of subspace subcodes of  optimum codes in rank metric}

\author{E.~M. Gabidulin\footnote{gab@pop3.mipt.ru} and P. Loidreau\footnote{Pierre.Loidreau@ensta.fr}}

\maketitle

\begin{abstract}
Maximum rank distance codes denoted MRD-codes are the equivalent in rank metric of 
MDS-codes. Given any integer $q$ power of a prime and any integer $n$  there is a family  
 of MRD-codes of length $n$ over $\FF{q^n}$ having  polynomial-time decoding 
algorithms. These codes can be seen as the analogs of Reed-Solomon codes 
(hereafter denoted RS-codes) for rank metric. 
In this paper their subspace subcodes are characterized. 
It is shown that hey are  equivalent to MRD-codes constructed in the same way 
but with smaller parameters. A specific polynomial-time decoding algorithm is designed.
Moreover, it is shown that the   direct sum of subspace subcodes  is 
 equivalent  to the direct product of MRD-codes with smaller parameters. This implies that
the decoding procedure can  correct errors of higher rank than the 
error-correcting capability. Finally it is shown that,  
for given parameters, subfield subcodes are completely characterized
by elements of the general linear group $\mbox{GL}_n(\FF{q})$ of non-singular $q$-ary matrices of size $n$.
\end{abstract}

\section{Introduction}

This work was initiated due to the 
great similarity between RS-codes and the family of   MRD-codes initially published  in \cite{Gabidulin:1985}.
It was also due  to the constant interest of research in the study of 
codes derived from RS-codes, particularly subfield subcodes of RS-codes and more recently subspace subcode
of RS-codes. Subspace subcodes or subgroup subcodes consist 
of the set of codewords whose components belong to a specific subspace or
subgroup of the additive group alphabet of the code \cite{Jensen:1995}. These particular subcodes are  in
general not linear but simply {\em additive}.

In the case of RS-codes abundant literature is available especially on subfield subcodes, that is when the considered
subspace is a subfield of the alphabet field. 
Indeed  RS-codes and derived families (GRS-codes for instance) form a very popular family of codes. 
Some  of the  most studied codes can be seen as particular subfield subcodes of  RS-codes or GRS-codes ( Alternant codes, binary  Goppa codes and BCH-codes for example ) \cite{Goppa:1970,MacWilliams/Sloane_Goppa:1977}. 
In cryptography also they play an important role, for instance in 
 McEliece public-key cryptosystem which uses in its design  the family of binary Goppa codes \cite{McEliece:1978}.

Concerning more general subcodes like subspace subcodes of RS-codes, research on the subject was initiated in 1992 \cite{Solomon:1992}. One of the  objects  of the research  is to provide longer character-oriented codes than RS-codes, that is to build codes with good parameters, but whose symbol length is controlled, smaller than the extension degree of the field.

In general however, even  the  simple question of the dimension of the subfield subcodes and their exact  minimum distance remains open, although some bounds are derived from bounds or equalities on parameters of  the parent RS-codes \cite{Delsarte:1975,Stichtenoth:1990,Bierbrauer/Edel:1997}. Concerning subspace subcodes, the same questions arise and 
again some bounds are obtained from RS-codes. More specifically,   it was
shown that their cardinality  depended on the structure of the subspace of the alphabet field,  relatively to 
the action of the Frobenius automorphism \cite{McEliece/Solomon:1994,Hattori/McEliece/Solomon:1998}.

An additional  problem the user of subspace subcode must cope with is the encoding of the code. Since
these codes  are not linear but simply additive, there is no generator matrix and therefore no systematic procedure can be built.  
In the case of {\em bit shortened RS-codes}  a systematic encoding procedure was designed \cite{Solomon:1992}. 
In a more general case of MDS codes an fast but not completely optimal  procedure was designed in \cite{Dijk/Tolhuizen:1999}.
 The principle consist of considering a codeword of length $n$ over $\FF{q^n}$ as an 
$m \times n$ $q$-ary matrix and by putting information on some subblock of the matrix, parities on some other block and some relations must be satisfied on the remaining positions.
However since  only  lower bound on the cardinality of the codes are known, these encoding procedures do not take into
account the additional bits that could be encoded and therefore are not optimal.

In the same way as RS-codes in Hamming metric the family of MRD-codes constructed in \cite{Gabidulin:1985} can be efficiently used for decoding in rank metric, for example,   whenever the errors
occur along some rows or columns of arrays, which  happens  along
tapes \cite{Roth:1991,Richter/Plass:2004_1}. Moreover, properties of rank metric have 
interesting cryptographic applications, in particular in the design of McEliece-like cryptosystems, see \cite{Gabidulin/Paramonov/Tretjakov:1991,Chen:1996}. 
 Namely, for the same set of parameters general purpose decoding algorithms in rank metric have a much higher
 complexity  compared to  general purpose decoding algorithms for Hamming metric \cite{Chabaud/Stern:1996,Ourivski/Johannson:2002,Canteaut/Chabaud:1998}. Therefore the MRD-codes or codes derived from MRD-codes can be of interest in designing cryptosystems. 
Several fast polynomial-time 
decoding algorithm up to the error-correction capability exist whose design that all have their equivalent in decoding  RS-codes. There are Euclidian 
and Berlekamp Massey like algorithms \cite{Gabidulin:1985,Gabidulin:1991,Roth:1991,Richter/Plass:2004_1} as well 
as Welch-Berlekamp like algorithms, \cite{Loidreau:2004,Loidreau:2005}.

 Let $\FF{q}$ be the base field and  $\FF{q^n}$ be an extension field of degree $n$ of $\FF{q}$. In the following we will indifferently consider $\FF{q^n}$ as the field or the $n$-dimensional vector space over $\FF{q}$. We recall properties of rank metric,  \cite{Gabidulin:1985}.

\begin{definition}[Rank of a vector]\label{Defi:Rang} \hfill

Let $\mathbf{e} = (e_1,\ldots,e_n) \in \FF{q^n}^n$. The rank over $\FF{q}$ of $\mathbf{e} $ is  
 the rank of the $n \times n$ $q$-ary matrix obtained by extending every component $e_i$ over a basis of 
$\FF{q^N}/\FF{q}$. It is denoted $\Rg(\mathbf{e}|\FF{q})$.
\end{definition}

The rank over $\FF{q}$ of  vector $\mathbf{e}$ is denoted in the following by $\Rg(\mathbf{e})$. 
We define $[i] \stackrel{def}{=} q^i$, when  $i\geq 0$ and $[i] \stackrel{def}{=} q^{n+i}$ when  $i< 0$.

A $[n,k,d]$-code over the field $\FF{q^n}$ has   
generator matrix 
\begin{equation}
\label{G2}{\mathbf G} = \left(
\begin{array}{lll}
g_1 &  \cdots  & g_n  \\
\vdots  & \ddots  &  \vdots  \\
g_1^{[k-1]} & \cdots  & g_n^{[k-1]}
\end{array}
\right) ,
\end{equation}
where $g_1  , \ldots  , g_n \in \FF{q^n}$ are linearly independent elements of   $\FF{q^n}/ \FF{q}$. 
A  parity-check matrix $\mathbf{H}$ of the code  has the same structure as $\mathbf{G}$ that is
 \begin{equation}
\label{H2}{\mathbf H=}\left(
\begin{array}{lll}
h_1 &  \cdots  & h_n \\
\vdots  & \ddots  &  \vdots  \\
h_1^{[d-2]} &  \cdots  & h_n^{[d-2]}
\end{array}
\right) ,
\end{equation}
for  $h_1 , \ldots  , h_n \in \FF{q^n}$  linearly independent over $\FF{q}$. The $h_i$'s of such codes can be easily 
found from the $g_i$'s \cite{Gabidulin:1985}.

The code $\mathcal{G}$ with parity-check matrix $\mathbf{H}$ or generator matrix $\mathbf{G}$ has 
minimum rank distance exactly  $d = n-k+1$. This code satisfies the Singleton like equality for rank metric, and
since Hamming metric is thinner than rank metric, this implies equally that it is a MDS code.  
There are several  polynomial-time decoding algorithms decoding these codes up to their error-correction capability \cite{Gabidulin:1985,Gabidulin:1991,Roth:1991,Richter/Plass:2004_1,Loidreau:2005}. It is remarkable to note that these algorithms are can be retranscripted from the algorithms decoding Reed-Solomon codes by replacing the notion of polynomial by the 
notion of linearized polynomials, that is inherent to the very definition of rank metric. Such similarity and such strong structure  raises the natural question of the structure of subcodes of MRD-codes.

The goal of this paper is to show that, despite the fact that the family of MRD-codes with parity-check matrix (\ref{H2}) are  very  similar to RS-codes, they
are much more structured, and all the question  concerning  the dimension of subspace subcodes, their exact minimum rank
distance and the design of  specific encoding-decoding procedures can be solved quite simply.

The paper is organized as follows: 
In a first part we show that subspace subcodes of a $(n,k = n-d+1 , d)$ MRD-code over  a $m$-dimensional subspace of $\FF{q^n}$
can be put in one-to-one correspondence with a $(m,k' = m-d+1 ,d)$ MRD-code over $\FF{q^n}$. More specifically we construct a 
bijective rank preserving $\FF{q}$-linear  mapping between the two codes. The mapping enables to build  specific encoding and decoding procedures.

In a second part we are interested in subcodes obtained from the direct sum of subspace subcodes 
and we show that it is possible to 
construct a rank-preserving mapping  putting  these subcodes in bijection 
with the direct product of MRD-codes. In that case we show that it is sometimes possible to correct beyond the
error-correcting capability of the codes.  

In a third part we deal with subfield subcodes of MRD-codes. We show that they 
are similar to the direct product of MRD-codes over the subfield, and that up to the 
action of the general linear group $\mbox{GL}_n(\FF{q})$  
on the components of the codewords, they can be uniquely defined. 
 This implies in particular that results from second part apply and that sometimes it is 
possible to decode them beyond the error-correcting capability of the codes.

\section{Subspace subcodes of rank codes}
\label{Section:Subcodes}

Let $\mathcal{G}$ be the code with generator matrix (\ref{G2}) and
parity-check matrix (\ref{H2}). Consider $V_m$ a $m$-dimensional subspace of $\FF{q^n}$.
Let 
\[
  ( \mathcal{G}| V_m) \stackrel{def}{=} \left\{  \mathbf{c} =(c_1,\ldots,c_n) \in \mathcal{G} ~|~ c_j\in V_m,~j=1,\ldots,n \right\}
\]

\begin{definition}\label{Defi:SuspaceSubcode} \hfill

  $(\mathcal{G} | V_m)$ is called subspace subcode of $\mathcal{G}$ over $V_m$.
\end{definition}

$(\mathcal{G} | V_m)$ is formed of the codewords whose components lie in the alphabet formed by the subspace $V_m$.
In a first section we construct a mapping between $( \mathcal{G}| V_m)$ and a so-called parent code $\mathcal{LG}(V_m)$.
This code is MRD and we show that the mapping is bijective, preserves $\FF{q}$-linearity and the rank. 
In a second part we describe encoding and decoding procedures for subspace subcodes.

\subsection{Characterization of subspace subcodes}

Let $\mathbf{c} =(c_1,\ldots,c_n),~ c_j\in V_m \mbox{ for all }j$.
Let   $\mathbf{b}=(\beta_1,\ldots,\beta_m)$  be a basis  of $V_m$.  
Vector  $\mathbf{c}$ has a unique decomposition  under the form 

\begin{equation}\label{cU}
        \mathbf c =\mathbf{b}U=(\beta_1,\ldots,\beta_m)U,
\end{equation}
where $U=(U_{ij})_{i=1,j=1}^{m,n} \in \FF{q}^{m \times n}$. 
Vector  $\mathbf{c}$ is a codeword if and only if it satisfies the parity-check equations

\begin{equation}\label{bUH}
        \mathbf{c} \mathbf{H}^T = (\beta_1,\ldots,\beta_m)U \mathbf{H}^T = \mathbf{0}.
\end{equation}

Hence $(\mathcal{G} | V_m)$ is  characterized  by the \emph{fixed} 
basis $ \mathbf{b} = (\beta_1,\ldots,\beta_m)$ and by the set of $m\times n$ matrices $U$ with coefficients in  $\FF{q}$ 
satisfying condition (\ref{bUH}). Solving (\ref{bUH}) is equivalent to solving 
\begin{equation}\label{bv}
        (\beta_1,\ldots,\beta_m)
\left(
\begin{array}{ccc}
        v_1 &  \cdots  & v_1^{[d-2]} \\
        \vdots  & \ddots & \vdots \\
        v_m  & \cdots & v_m^{[d-2]} 
\end{array}
\right)
=\mathbf{0},
\end{equation}
where 
\begin{equation}\label{Eq:V}
(v_1,\ldots,v_m) = (h_1,\ldots,h_n) U^t.
\end{equation}
Given any vector $(v_1,\ldots,v_m) \in \FF{q^n}^m$, there exists a unique $m \times n$ $q$-ary matrix $U$ 
such that 
 $(v_1,\ldots,v_m) = (h_1,\ldots,h_n) U^t$. The $i$th column of $U$ is given  by the vector obtained from   the  representation 
of $v_i$  over the basis  $(h_1,\ldots,h_n)$.

 Condition (\ref{bv}) is equivalent  to 
\begin{equation}\label{vb}
(v_1,\ldots,v_m)
\left(
\begin{array}{ccc}
        \beta_1^{[n]} & \ldots & \beta_1^{[n-d+2]} \\
        \vdots &  \ddots & \vdots \\
        \beta_m^{[n]} & \ldots & \beta_m^{[n-d+2]}\\
\end{array}
\right)
=\mathbf{0}.
\end{equation}

Since $\beta_1,\ldots, \beta_m$ are linearly independent, equation (\ref{vb}) implies 
that $\mathbf{v}=(v_1,\ldots,v_m)$ is a codeword of  a $\FF{q^n}$-linear MRD-code 
with parameters  $[m,m-d+1,d]$. The code with parity-check matrix

\begin{equation}\label{HVm}
        \mathbf{H}_{V_m}=
        \left(
        \begin{array}{ccc}
        \beta_1^{[n]} & \cdots & \beta_m^{[n]}  \\
        \vdots  & \ddots & \vdots \\
        \beta_1^{[n-d+2]} & \cdots & \beta_m^{[n-d+2]}\\
\end{array}
\right)
\end{equation}
is  denoted  $\mathcal{LG}(V_m)$

\begin{definition}[Parent Code] \label{Defi:ParentCode} \hfill

The code  $\mathcal{LG}(V_m)$  is called  the parent code of $(\mathcal{G}|V_m)$. 
\end{definition}

We now prove  the following proposition establishing that any 
subspace subcode of  a MRD-code of full length is uniquely characterized by a MRD-code 
with the same minimum distance but with smaller parameters.  

\begin{prop}\label{Prop:Bijection} \hfill

   Let $\mathbf{b} = ( \beta_1,\ldots,\beta_m ) $ be a basis of $V_m$
   over $\FF{q}$, and let $\mathbf{h} = (h_1,\ldots,h_n)$ be the vector
   defined in equation (\ref{H2}). The mapping
\[
\begin{array}{lcl}
  f_{\mathbf{b}} : ~V_m^n   & \rightarrow &\FF{q^n}^m \\
  \mathbf{c} = \mathbf{b}U  & \mapsto & f_{\mathbf{b}}( \mathbf{c} ) = \mathbf{h} U^t
\end{array}
\]
satisfies the following properties
\begin{enumerate}
\item $f_{\mathbf{b}}$ is a $\FF{q}$-linear bijective mapping. 
\item $f_{\mathbf{b}}$ preserves the rank of vectors over $\FF{q}$, that is $\Rg \left( f_{\mathbf{b}}(\mathbf{c}) | \FF{q} \right)  = \Rg(\mathbf{c}| \FF{q})$.
\item $f_{\mathbf{b}} \left( \mathcal{G}|V_m \right) = \mathcal{LG}(V_m)$.
\item $f_{\mathbf{b}}$ and $f_{\mathbf{b}}^{-1}$ can be computed in $nm$ multiplications in $\FF{q}$ and
  $n$ additions in $\FF{q^n}$.  
\end{enumerate}
\end{prop}

\begin{proof}

  \begin{enumerate}
  \item Since $h_1,\ldots,h_n$ are linearly independent over $\FF{q}$, it follows that $f_{\mathbf{b}}$ is 
a bijection. Let  $\mathbf{c} = \mathbf{b}U$ and   $ \mathbf{d} = \mathbf{b}V$
be vectors of $V_m^n$. By definition of   $f_{\mathbf{b}}$, we have  
$f_{\mathbf{b}}(\mathbf{c} + \mathbf{d}) = \mathbf{b} (U+V) =   f_{\mathbf{b}}(\mathbf{c}) + f_{\mathbf{b}}(\mathbf{d})$.

  \item Let $\mathbf{c} = \mathbf{b}U$. Since  $\beta_1,\ldots,\beta_m$ are
    linearly independent, we have that $\Rg( \mathbf{c} | \FF{q}) = \Rg(U)$, where
    $\Rg(U)$ is the rank of  matrix $U$. Moreover, since  $f_{\mathbf{b}}(\mathbf{c}) = \mathbf{h}U^t$,
   and $h_1,\ldots,h_n$ are linearly independent,we have that 
    \[
    \Rg( f_{\mathbf{b}}(\mathbf{c}) | \FF{q}) = \Rg(U^t) = \Rg(U) =  \Rg( \mathbf{c} | \FF{q})
    \]

\item The fact that $f_{\mathbf{b}} \left( \mathcal{G}|V_m \right) = \mathcal{LG}(V_m)$,
  follows directly from the definition of mapping $f_{\mathbf{b}}$. 

\item Any vector of  $V_m^n$ is given by a $q$-ary $m \times n$ matrix $U$. Therefore, computing $f_{\mathbf{b}}$ is
merely computing the product of vector  $\mathbf{h} = (h_1,\ldots,h_n)$ by matrix $U$. This can be done in $nm$ multiplications in $\FF{q}$ and $n$ additions in $\FF{q^m}$. Conversely, computing $f_{\mathbf{b}}^{-1}(v_1,\ldots,v_m)$ corresponds to finding the unique $q$-ary matrix $U$ such that $ (v_1,\ldots,v_m) = (h_1,\ldots,h_n) U^t$, and then compute $\mathbf{b}U$.

\end{enumerate}
\end{proof}

\noindent We deduce the following  corollary. 
\begin{corollary}\label{Cor:SousCode} \hfill

  $\left( \mathcal{G}|V_m \right)$ is a $(n,M,D)$-additive code, where 
  \begin{itemize}
  \item $D = d$, 
  \item $M = q^{n(m-D+1)}$.
  \end{itemize}

\end{corollary}

Both statements of the corollary
 imply that  $\left( \mathcal{G}|V_m \right)$  is optimal for the rank metric \cite{OGHA:2003}.
  
\subsection{Coding and decoding of subspace subcodes}
\label{Section:Coding}

Thanks to the mapping $f_{\mathbf{b}}$ described in proposition \ref{Prop:Bijection}, we design an efficient 
encoding procedure  for subspace subcodes of MRD-codes.  From Corollary \ref{Cor:SousCode} the number of 
$q$-ary digits that can be encoded is equal to $n(m-d+1)$. Hence any information vector can
be considered as a vector of length  $(m-d+1)$ over $\FF{q^n}$.

Let $\mathbf{x} = (x_1,\ldots, x_{m-d+1}) \in \FF{q^n}^{m-d+1}$ be  an  information vector. 
Let $\mathbf{G}_{V_m}$ be a generator matrix of the parent code $\mathcal{LG}(V_m)$.

The encoding procedure is the following: 
        \begin{enumerate}
                \item Encoding in the parent code: Compute $\mathbf{y} = \mathbf{x}\mathbf{G}_.{V_m} \in \mathcal{LG}(V_m)$; 
                \item Transferring in the subcode: Compute $\mathbf{c} = f_{\mathbf{b}} ^{-1}(\mathbf{y})$.    
        \end{enumerate}

The complexity of the encoding  procedure  consists of  $(m-d+1)m $ multiplications in $\FF{q^n}$ 
if we neglect 
the operations over the base field $\FF{q}$.

Let   $\mathbf{y} = \mathbf{c} + \mathbf{e}$ be a received vector where $\mathbf{c} \in 
(\mathcal{G} | V_m)$ and $\mathbf{e}$ has coefficients in  $V_m$  and  rank $t \le \lfloor (d-1)/2 \rfloor$. 
There are two manners of decoding:

        \begin{itemize}
        \item In the code $\mathcal{G}$  by  using the standard decoding algorithms for $\mathcal{G}$. 
The complexity is $\approx (d-1+t)n  + t^3$  multiplications in $\FF{q^n}$ if we take the 
decoding algorithm described in \cite{Gabidulin:1991}.
        
        \item By  decoding in  the parent code $\mathcal{LG}(V_m)$: We have  $f_{\mathbf{b}} (\mathbf{y}) = f_{\mathbf{b}}(\mathbf{c}) + f_{\mathbf{b}} (\mathbf{e}) $, where $f_{\mathbf{b}}(\mathbf{c}) \in \mathcal{LG}(V_m)$, and $\Rg(f_{\mathbf{b}}(\mathbf{e}) | \FF{q}) = t$. Therefore, by correcting $f_{\mathbf{b}}(\mathbf{y})$ in $\mathcal{LG}(V_m)$, one recovers
$f_{\mathbf{b}}(\mathbf{c})$ and $f_{\mathbf{b}}(\mathbf{e})$ and by computing the inverse function one gets 
$\mathbf{c}$ and $\mathbf{e}$. 
The complexity of the algorithm is
$\approx (d-1)m + tn +t^3$ multiplications in $\FF{q^n}$. 
        
\end{itemize}

\subsection{Direct sum of subspace subcodes}

In the previous section, we showed that subspace subcodes of MRD-codes are in 
some sense isomorphic  to MRD-codes of smaller length. 
From subspace
subcodes, we build codes corresponding to the direct sum of subspace
subcodes. The mapping $f_{\mathbf{b}}$ can be extended to this direct sum.

Consider a sequence $V_{m_1},\ldots,V_{m_u}$ of subspaces
of $\FF{q^m}$ of dimensions $m_i$ that two-by-two do not intersect except on the zero-vector (this implies in particular that $\sum_{i=1}^u{m_i} \le n$). For
every subspace $V_{m_i}$ we fix a basis  $\mathbf{b}_i$. 

As before $( \mathcal{G} | V_{m_i})$ denotes the  subspace subcode of the code $\mathcal{G}$
restricted to vectors with coordinates in $V_{m_i}$. Let 
\[
\mathcal{M} \stackrel{def}{=} ( \mathcal{G} | V_{m_1}) \oplus \cdots \oplus ( \mathcal{G} |
V_{m_u} )\subset \mathcal{G}
\]
be the subcode of $\mathcal{G}$ consisting of the direct sum of the subspace
subcodes $( \mathcal{G} | V_{m_i})$, that is    

\begin{equation}\label{decomp1}
        \mathcal{M}  =\left\{\mathbf{c}=\mathbf{c}_1+ \cdots + \mathbf{c}_u
        ~|~ \mathbf{c}_1 \in (\mathcal{G}| V_{m_1}),\ldots,\mathbf{c}_u \in
        (\mathcal{G} | V_{m_u} )\right\}.
\end{equation}
 We define the mapping $f_{( \mathbf{b}_1,\ldots,\mathbf{b}_u)} $ from restricted mappings $f_{\mathbf{b}_i}$ as defined
in proposition \ref{Prop:Bijection}: 

\[
\begin{array}{lcl}
  V_{m_1}^n \oplus \cdots \oplus  V_{m_u}^n  & \rightarrow &  \FF{q^n}^{m_1}
  \times \cdots \FF{q^n}^{m_u} \\
  \mathbf{c} = \mathbf{c}_1 + \cdots + \mathbf{c}_u & \mapsto & f_{( \mathbf{b}_1,\ldots,\mathbf{b}_u)}( \mathbf{c}) = \left(
  f_{\mathbf{b}_1}(\mathbf{c}_1),\ldots,f_{\mathbf{b}_u}(\mathbf{c}_u) \right)
\end{array}
\]

 \begin{prop} \hfill
   
   $ f_{ ( \mathbf{b}_1, \ldots, \mathbf{b}_u ) }$ is  $\FF{q}$-linear, bijective and preserves the rank. 

\end{prop}

\begin{proof}

  $\FF{q}$-linearity and bijectivity come from the fact that $ f_{( \mathbf{b}_1,\ldots,\mathbf{b}_u )}$  is a direct product of $\FF{q}$-linear, bijective mappings.

Concerning the rank property, any vector   $\mathbf{c} \in V_{m_1}^n \oplus \cdots \oplus  V_{m_u}^n$  can be uniquely written under the form 
\[
\mathbf{c} = \mathbf{b}_1 U_1 + \cdots + \mathbf{b}_u U_u,
\]
where $U_i$'s are $n m_i$ $q$-ary matrices. This can be rewritten under the form 
\[
\mathbf{c} = (\mathbf{b}_1,\ldots,\mathbf{b}_u)
\left(
\begin{array}{c}
  U_1 \\
  \vdots \\
  U_u
\end{array}
\right).
\]
Since, for $i=1,\ldots,u$ vector  spaces $V_i$ form a direct sum, this implies that the components of vector $(\mathbf{b}_1,\ldots,\mathbf{b}_u)$ are linearly independent over $\FF{q}$. Therefore the rank of $\mathbf{c}$ over $\FF{q}$ is equal to the rank of 
matrix
\[ \mathcal{U} = 
\left(
\begin{array}{c}
  U_1 \\
  \vdots \\
  U_u
\end{array}
\right).
\]
Therefore, since  $f_{( \mathbf{b}_1,\ldots,\mathbf{b}_u)}( \mathbf{c}) = \mathbf{h} \mathcal{U}^t$,  we have 
\[
 \Rg \left(  f_{( \mathbf{b}_1,\ldots,\mathbf{b}_u)}( \mathbf{c}) | \FF{q} \right) =  \Rg(\mathbf{c} | \FF{q}).
\]
\end{proof}

Let   $\mathcal{LG}(\mathcal{M})$ be  the code with parity-check matrix, 
\begin{equation}\label{HM}
\mathbf{H}(\mathcal M)= \left[\begin{array}{lll}
        \mathbf{H}_{V_{m_1}} & \cdots  &  0 \\
        \vdots & \ddots & \vdots \\
         0   & \cdots & \mathbf{H}_{V_{m_u}}
        \end{array}\right].
\end{equation}

We prove the following proposition
\begin{prop} 
$ f_{( \mathbf{b}_1,\ldots,\mathbf{b}_u ) }(\mathcal{M})  = \mathcal{LG}(\mathcal M)$.
\end{prop}

\begin{proof} \hfill

Let $\mathbf{H}$ be the parity-check matrix of $\mathcal{G}$ under the form (\ref{H2}).
  Let $\mathbf{c} = \mathbf{c}_1 + \cdots + \mathbf{c}_u \in \mathcal{M}$. Since for all
$i=1,\ldots,u, ~ \mathbf{c}_i \in (\mathcal{G}| V_{m_i})$ we have for all $i=1,\ldots,u, ~ \mathbf{c}_i \mathbf{H} = 0$. This is equivalent to 
 $f_{\mathbf{b}_i}(\mathbf{c}_i) \mathbf{H}_{V_{m_i}} = 0$, for all $i=1,\ldots,u$.

\end{proof}

For this reason  $\mathcal{LG}(\mathcal M)$ is called parent code of
$\mathcal{M}$. As it is also a  direct product of 
MRD-codes with smaller parameters, we deduce  the following corollary. 

\begin{corollary}\label{Prop:DirectSum} 
  $\mathcal{M}$ is a $(n,M,D)$-code, where
\begin{itemize}
\item $M = q^{n \sum_{i=1}^u{ ( m_i -(d-1) ) } }$.
\item $D = d$.
\end{itemize}
\end{corollary}

From $f_{( \mathbf{b}_1,\ldots,\mathbf{b}_u )}$ we deduce   
efficient specific encoding and decoding procedures for code $\mathcal{M}$.

Let $\mathbf{x}$ be $q^n$-ary vector of length  $ \sum_{i=1}^u{m_i} -u(d-1)$. 
\begin{enumerate}
\item Vector $\mathbf{x}$ is first  divided into $u$ blocks $\mathbf{x}_i$ each of length $ m_i - d + 1 $. 
\item Any subvector $\mathbf{x}_i$ is encoded into $\mathbf{c}_i \in ( \mathcal{G}~|~ V_{m_i}) $, using the procedure described
in  section \ref{Section:Coding}  with mapping $f_{ \mathbf{b}_i}$.  
\item The encoded codeword is $\mathbf{c} =\mathbf{c}_1 + \cdots + \mathbf{c}_u  \in \mathcal{M}$.
\end{enumerate}

Since $\mathcal{LG}(V_m)$ can be decomposed into a direct product of subspace subcodes, 
we show that we can go  further in the decoding
and  that  it is sometimes possible to decode beyond the
error-correcting capability  $C \stackrel{def}{=} \lfloor (d-1)/2 \rfloor$
of the code.  Let  the received vector 
\[
        \mathbf{y}  = \mathbf{c} + \mathbf{e} \in   V_{m_1}^n \oplus \cdots \oplus  V_{m_u}^n,
\]  
 where $\mathbf{c} \in \mathcal{M}$ and $\mathbf{e}$ is some error-vector
 of rank less than $t$. Let $\mathbf{y}_i$ be the projection of $\mathbf{y}$ on subspace $V_{m_i}$. We
 have the following set of equations 
\[
\left\{ 
\begin{array}{l}
        \mathbf{y}_1 = \mathbf{c}_1 + \mathbf{e}_1, ~\mathbf{c}_1 \in
        (\mathcal{G} | V_{m_1}), \\
        \vdots \\ 
        \mathbf{y}_u = \mathbf{c}_u + \mathbf{e}_u,~\mathbf{c}_u \in (\mathcal{G}| V_{m_u}).
\end{array}
\right.
\]
where, for all $i=1,\ldots,u$ the rank of  $\mathbf{e}_i$  over $\FF{q}$ is less or equal to $t$. 
 Therefore, if  $t \le C$   the $\mathbf{y}_i$'s  are decodable in their respective subcodes $(\mathcal{G}| V_{m_i})$. 
Hence  $\mathbf{y}$ is decodable in $\mathcal{M}$. Moreover,  even when  $\Rg(\mathbf{e}) > C$, 
it is sometimes possible to decode successfully. This corresponds to the case where $\Rg(\mathbf{e}_i)
 \le C$ for all $i=1,\ldots,u$.

We study  occurrences of such case. We want to find an estimation of: 
\[
   P_{decoding} =  Pr( \Rg(\mathbf{e_1})\le C,\ldots, \Rg(\mathbf{e_u}) \le C) ~|~ \Rg(\mathbf{e}) \le t), 
\]
which quantifies the probability of successful decoding in  $\mathcal{M}$. 

Let  $N = \sum_{i=1}^u{m_i}$, and let us  consider  
the error-vector $\mathbf{e}$ as the $q$-ary $N \times n$  
matrix corresponding to the expansion rowwise of the components of  $\mathbf{e}$ on 
the basis  of the $N$ dimensional vector-space $  V_{m_1} \oplus \cdots \oplus  V_{m_u}$ with basis $(\mathbf{b}_1,\ldots,\mathbf{b}_u)$. This gives the following representation for $\mathbf{e}$
\[
\mathbf{e} = \left( 
\begin{array}{c}
  \mathbf{e_1} \\
  \vdots \\
\mathbf{e_u}
\end{array}
\right)
\begin{array}{c}
  \mathbf{b}_1 \\
  \vdots \\
  \mathbf{b}_u
\end{array}
\]
where $\mathbf{e}_i$'s are $q$-ary matrices of size $m_i \times n$.
We suppose that  $\mathbf{e}^T $ is of rank $ \le t$, where $t$ is some
integer.  

Matrix  $\mathbf{e}^T$  can be transformed into a vector 
$\mathbf{E} = (\mathbf{E}_1 \cdots \mathbf{E}_u)$ with components in $\FF{q^n}$ by 
considering  any of its columns  as the coordinates of an element of $\FF{q^n}$ on 
a basis of $\FF{q^n}/\FF{q}$. This implies that $\Rg(\mathbf{E}) \le t$ if and only 
if $\Rg(\mathbf{e}) \le t$. In particular there exist   $\alpha_1,\ldots,\alpha_t \in \FF{q^n}^t$ 
linearly independent over $\FF{q}$  satisfying 
\[
 \mathbf{E} = (\alpha_1,\ldots,\alpha_t) S,
\]
where $S$ is a $t \times N$-matrix over $\FF{q}$ of rank less than $t$. 
Consider the decomposition of $S = (S_1 \cdots S_u)$, where $S_i$ are $t \times m_i$ $q$-ary matrices. We have 
\[
\left\{
\begin{array}{l}
        \mathbf{E}_1 =   (\alpha_1,\ldots,\alpha_t) S_1, \\
        \vdots  \\
        \mathbf{E}_u  =  (\alpha_1,\ldots,\alpha_t) S_u.
\end{array}
\right.
\]
Since the transformations from $\mathbf{e}_i$  to $\mathbf{E}_i$ are one-to-one and preserve the rank, and 
since the rank of $  \mathbf{E}_i$  over $\FF{q}$ is equal to the rank of $S_i$ we have 
\[  
P_{decoding} =  Pr( \Rg(S_1) \le C,\ldots, \Rg(S_u) \le C) ~|~ \Rg(S) \le t  ).
\]
Matrix $S$  being of size  $t \times n$, the conditioning  on the rank of $S$ is always satisfied.
Therefore we can remove it and we obtain 
\[  
P_{decoding} =  Pr( \Rg(S_1) \le C,\ldots, \Rg(S_u) \le C) ).
\]
The events being independent, this is equivalent to  
\begin{equation}\label{Eq:Proba}
   P_{decoding} =  Pr( \Rg(S_1) \le C ) \cdots  Pr( \Rg(S_u) \le C). 
\end{equation}

 The  number $\mathcal{N}_C(m,t) $ of $t \times m $  $q$-ary matrices of rank $C$ 
is given by the formula, see \cite{Lidl/Niederreiter:1997} page $455 $
for instance: 
\[
        \mathcal{N}_C(m,t) =  \prod_{i=0}^{C-1}{\frac{(q^m - q^i)(q^t-q^i)}{q^C-q^i}}.
\]
This quantity  can be  approximated by  $q^{(m+t)C -C^2 + q^{-1} + O( q^{-2})}$, Therefore 
\begin{equation}\label{Eq:ProbaPartielle} 
Pr( \Rg(S_i) \le C ) =  q^{ ( m_i - C ) ( C - t)   + q^{-1} + O( q^{-2} ) },
\end{equation}

 Hence from (\ref{Eq:Proba})  and (\ref{Eq:ProbaPartielle}) we obtain 

\begin{prop}[Probability of successful decoding]\label{Prop:Decoding} \hfill

Let $\mathcal{M} = ( \mathcal{G} | V_{m_1}) \oplus \cdots \oplus ( \mathcal{G} | V_{m_u} )$,
be the code formed by the direct sum of subspace subcodes of maximum rank distance codes with 
error-correcting capability $C$.    
Then the probability of success for decoding $t > C$ errors satisfies 
\[
P_{decoding} =  q^{  - ( N - C ) (t - C ) + u q^{-1} + O( q^{-2} )  },  
\]
where $N = \sum_{i=1}^u{m_i}$.
\end{prop}

\section{A particular case: subfield subcodes}

Subfield subcodes are special  cases of subspace subcodes. In Hamming metric  constructing 
a parity-check or   generator matrix for these codes by using properties of the Trace operator is easy, \cite{Delsarte:1975,Goppa:1970,MacWilliams/Sloane_Goppa:1977}. What is less trivial is computing the exact dimension and exact minimum 
distance of the codes. Generally speaking, only bounds are available.

Section \ref{Section:Subcodes} showed that the exact parameters could be obtained very 
simply for the family of MRD-codes with parity-check matrix \ref{H2}.
Namely, subspace subcodes are isomorphic through a rank preserving bijection to  MRD-codes of the same family but smaller
parameters. This implies in  particular that subspace subcodes are optimal for rank metric.

In this section we go one step further and show that, given a subfield $\FF{q^s}$ of $\FF{q^n}$ specified 
by a chosen basis, there is a unique subfield subcode  modulo transformation by the group 
induced on the components of the code by the general linear group $GL_n(\FF{q})$ of
of $q$-ary non-singular matrices of size $n \times n$. We prove the following theorem.

\begin{theorem}\hfill
     
 Let $\mathcal{G}$ be a code over $\FF{q^n}$ with parity-check matrix (\ref{H2}), 
Let $s$ be a positive integer dividing $n$ and let 
\[
A =     \left(
        \begin{array}{ccc}
        a_1 & \cdots & a_{s}  \\
        \vdots  & \ddots & \vdots \\
        a_1^{[d-2]} & \cdots & a_{s}^{[d-2]} 
\end{array}
\right) .
\]
where the $a_i \in  \FF{q^{s}} \subset \FF{q^n}$ for all $i=1,\ldots, s$  are  linearly independent over $\FF{q}$.

Then,  there exists a {\em unique} matrix $S \in \mbox{GL}_n(\FF{q})$ of size $n \times n$
such that the subfield subcode $( \mathcal{G} | \FF{q^s})$ has  parity-check matrix
\[
    \mathbf{H}_{q^s} =    
\left(
\begin{array}{cccc}
  A & 0 & \cdots & 0 \\
  0 & A & \cdots & 0 \\
  \vdots & \vdots & \ddots & \vdots \\
  0 & 0 & \cdots & A 
\end{array}
\right)
    S,
\] 
\end{theorem}

\begin{proof} \hfill

$\mathcal{G}$ has parity-check matrix (\ref{H2}) which can be rewritten 
  \[
  \mathbf{H} = \left( 
\begin{array}{c}
  \mathbf{h} \\
  \vdots \\
  \mathbf{h}^{[d-2]}
\end{array}
\right)
  \]
where $\mathbf{h}^{[i]} \stackrel{def}{=} \left( h_1^{[i]},\ldots,h_n^{[i]} \right)$.
One obtains a parity-check matrix of  $( \mathcal{G} | \FF{q^s})$ by the  following procedure:

\begin{itemize}
\item Choose a basis of $\FF{q^n}/ \FF{q^{s}}$.

\item Expand each line of matrix $\mathbf{H}$ 
onto $\FF{q^{s}}$ with respect to this basis. 
Every line of length $n$ with coefficients in $\FF{q^n}$ 
is transformed columnwise into a matrix of size $n/s \times n$, that is 
\[
\mathbf{h} = (h_1,\ldots,h_n) \mapsto 
 \mathcal{H} = \left( 
\begin{array}{ccc}
  h_{1,1} & \cdots & h_{1,n} \\
  \vdots & \ddots & \vdots \\
   h_{n/s,1} & \cdots & h_{n/s,n}
\end{array}
\right);
\]
\end{itemize}
However, since $\mathbf{H}$ is composed of  lines  obtained by the action of powers of the Frobenius automorphism on the components of   $\mathbf{h}$, for all $i=1,\ldots, d-2$, there exists a $n/s \times n/s$ non-singular matrix $Q_i$ with coefficients in $\FF{q^s}$ satisfying  

\[  
  \mathbf{h}^{[i]} = (h_1^{[i]},\ldots,h_n^{[i]}) \mapsto Q_i \mathcal{H}^{[i]},   
\]
where $\mathcal{H}^{[i]}$ denotes matrix  $\mathcal{H}$ whose components have been elevated to the power
$q^i$. Therefore, a parity-check matrix of  $( \mathcal{G} | \FF{q^s})$ has the form
\[
 \mathbf{H}_{q^s} = 
 \left( 
\begin{array}{c}
   \mathcal{H} \\
     \mathcal{H}^{[1]}\\
     \vdots \\
     \mathcal{H}^{[d-2]} 
\end{array}
\right)
\]
 Next, since the columns of $\mathcal{H}$ have rank $n$ over $\FF{q}$ (the $h_i$s are linearly independent over $\FF{q}$), there is a $n \times n$ matrix $S$ with coefficients in $\FF{q}$ such that
\[
\mathcal{H}  S = 
\left(
\begin{array}{cccc}
\mathbf{a}_1 & 0 & \cdots & 0 \\
0 &  \mathbf{a}_2 & \cdots & 0 \\
\vdots & \cdots & \ddots & \vdots \\
0 & 0 & \cdots &  \mathbf{a}_{n/s} 
\end{array}
\right),
\]
where, for all $i=0,\ldots,d-2$ 
\[
\mathbf{a}_i = (a_{i,1}, \ldots, a_{i,s}), \quad a_{i,j} \in \FF{q^{s}} 
\]
is a vector formed with linearly independent elements of $\FF{q^{s}}$. Provided 
$d-2 < s$ the following matrices  
\[
  \mathcal{A}_i = 
\left(
\begin{array}{c}
\mathbf{a}_i \\ 
\vdots \\
\mathbf{a}_i^{[d-2]}
\end{array}
\right), \quad \mbox{for } i=1,\ldots,n/s
\]
are generator matrices of a $[s,s-d+1,d]$ MRD-code over $\FF{q^s}$. 
This implies the existence of  a permutation matrix $P$ such that 
\[
\mathbf{H}_{q^s} = 
 \left( 
\begin{array}{c}
   \mathcal{H} \\
     \mathcal{H}^{[1]}\\
     \vdots \\
     \mathcal{H}^{[d-2]} 
\end{array}
\right)
= 
P
 \left( 
\begin{array}{ccc}
   \mathcal{A}_1 & \cdots  & 0\\
    \vdots & \ddots & \vdots \\
    0 & \cdots &  \mathcal{A}_{n/s}
\end{array}
\right)
S^{-1}
\]
To complete the proof we have  to remark that, multiplying a parity-check matrix on the left by any non-singular matrix 
doesn't change the generated code.  Hence  a parity-check of $(\mathcal{G}| \FF{q^s})$ is 
\[
\left( 
\begin{array}{ccc}
   \mathcal{A}_1 & \cdots  & 0\\
    \vdots & \ddots & \vdots \\
    0 & \cdots &  \mathcal{A}_{n/s}
\end{array}
\right)
S
\]
If   $A \stackrel{def}{=} \mathcal{A}_1$,  for all $i=1,\ldots,n/s$ there exists 
a non-singular  matrix $P_i$ over $\FF{q^s}$ such that $\mathcal{A}_i  = P_i \mathcal{A} $. Hence 
\[
\left( 
\begin{array}{ccc}
   \mathcal{A}_1 & \cdots  & 0\\
    \vdots & \ddots & \vdots \\
    0 & \cdots &  \mathcal{A}_{n/s}
\end{array}
\right)
 =  \left( 
\begin{array}{ccc}
   P_1 & \cdots  & 0\\
    \vdots & \ddots & \vdots \\
    0 & \cdots &  P_{n/s}
\end{array}
\right)
\left( 
\begin{array}{ccc}
   A & \cdots  & 0\\
    \vdots & \ddots & \vdots \\
    0 & \cdots &  A
\end{array}
\right), 
\]
and a parity-check matrix of $(\mathcal{C}| \FF{q^s})$ is 
\[
\mathbf{H}_{q^s} = 
 \left( 
\begin{array}{ccc}
   A & \cdots  & 0\\
    \vdots & \ddots & \vdots \\
    0 & \cdots &  A
\end{array}
\right) S. 
\]
This completes the proof.
\end{proof}

The theorem means that, somehow, the subfield subcode of a maximum rank distance code of full length (\emph{i.e.} 
the length of the code is equal to the extension degree) is a  direct sum of maximum rank 
distance codes taken over the  subfield. 

It also implies that, whatever the MRD-code over $\FF{q^n}$ be, if 
we fix a basis of $\FF{q^n}/\FF{q^s}$, then the subfield subcode is uniquely determined by a $q$-ary invertible matrix $S$.

From proposition \ref{Prop:Decoding} we deduce the following corollary 
\begin{corollary}[Successful decoding of subfield subcodes] \hfill

  Let $C$ be the error-correcting capability of $\mathcal{G}$, 
  then the probability of decoding $t > C$  errors in $  ( \mathcal{G} | \FF{q^s})$ is equal to 
\[
P_{decoding} =  q^{ - ( n - C ) (t - C) + \frac{n}{s}  q^{-1} + O( q^{-2} )  },  
\]
\end{corollary}

\end{document}